\documentclass[twocolumn,prl,aps,showpacs,superscriptaddress,floatfix]{revtex4}

\usepackage[dvips]{color,graphicx}
\usepackage{dcolumn}
\usepackage{bm}
\usepackage{latexsym}
\usepackage{amssymb}

\begin{document}

\title{Strongly correlated fermions after a quantum quench}

\author{S.R.\ Manmana}
\affiliation{Institut f\"ur Theoretische Physik III, Universit\"at
  Stuttgart, Pfaffenwaldring 57, D-70550 Stuttgart, Germany.} 
\affiliation{Fachbereich Physik, Philipps-Universit\"at Marburg,
  D-35032 Marburg, Germany.}
\author {S.\ Wessel} 
\affiliation{Institut f\"ur Theoretische Physik III, Universit\"at
  Stuttgart, Pfaffenwaldring 57, D-70550 Stuttgart, Germany.} 
\author{R.M.\ Noack} 
\affiliation{Fachbereich Physik, Philipps-Universit\"at Marburg,
  D-35032 Marburg, Germany.} 
\author{A.\ Muramatsu}
\affiliation{Institut f\"ur Theoretische Physik III, Universit\"at
  Stuttgart, Pfaffenwaldring 57, D-70550 Stuttgart, Germany.} 

\date{\today}

\pacs{05.30.-d, 05.30.Fk, 71.10.-w, 71.10.Fd}

\begin{abstract}
Using the adaptive time-dependent density-matrix renormalization group
method, we study the time evolution of strongly
correlated spinless fermions on a one-dimensional lattice after 
a sudden change of the interaction strength.
For certain parameter values, two different initial states
(e.g., metallic and insulating), 
lead to observables which become
indistinguishable after relaxation. 
We find that the resulting quasi-stationary state is non-thermal.
This result holds for both integrable and non-integrable
variants of the system.
\end{abstract}

\maketitle

Recent experiments on optical lattices have made it possible to
investigate the behavior of strongly correlated quantum systems 
after they have been quenched.
In these experiments, the system is prepared in an initial state
$|\psi_0 \rangle$ and then is pushed out of equilibrium by suddenly
changing one of the parameters. 
Prominent examples are the collapse and revival of a Bose-Einstein
condensate (BEC) \cite{Nature:Bloch}, the realization of a quantum
version of Newton's cradle \cite{Nature:Weiss}, and the quenching of a
ferromagnetic spinor BEC \cite{Nature:Sadler}. 
All of these systems can be considered to be closed, i.e., have no
significant exchange of energy with a heat bath, so that energy is conserved to
a very good approximation during the time evolution. 
Furthermore, since these systems are characterized by a large number
of interacting degrees of freedom,
application of the ergodic hypothesis leads to the expectation that
the time average of observables should become equal to thermal
averages after sufficiently long times. 
Various authors have recently given voice to such an expectation
\cite{sengupta:053616,berges:142002,corinna:quench}.
However, the experiment on one dimensional (1D) interacting bosons
shows no thermalization, a behavior that was
ascribed to integrability \cite{Nature:Weiss}. 
Rigol {\em et al.}\ found that an integrable system of hard-core
bosons relaxes to a state
well-described by a Gibbs ensemble that takes into account the full set of
constants of motion \cite{marcos:relax}; similar results were found for
the integrable Luttinger model \cite{cazalilla:156403}.

For a closed system, the set of expectation values of all powers of
the Hamiltonian $\hat H$ constitute an infinite number of constants of
motion, irrespective of its integrability.
Therefore, the question of the importance of integrability
in a closed system that is quenched arises.
We address this issue by investigating the full time evolution of a
strongly correlated system whose integrability can be easily destroyed 
by turning on an additional interaction term. 
We show, using the recently developed adaptive time-dependent density
matrix renormalization group method (t-DMRG) 
\cite{Uli_timeevolve,white:076401,luo:049701,schmitteckert:121302,manmana:269}, 
that, in a certain parameter range, two different initial states with
the same energy relax, to within numerical precision, to states with indistinguishable
momentum distribution functions.
A comparison with quantum Monte Carlo (QMC) simulations shows,
however, that they do not correspond to a thermal state.
By using a generalized Gibbs ensemble
\cite{balian,marcos:relax,cazalilla:156403,marcos:relax2} with the
expectation value of the powers of the Hamiltonian
$\langle {\hat H}^n \rangle$ as constraints, we can improve the agreement 
with the time averages of the evolved system.
This applies to both the integrable as well as the
non-integrable case.

In this Letter, we investigate the Hamiltonian  
\begin{equation}
\hat{H} = - t_{\rm h} \sum_j \left( c_{j+1}^{\dagger}
  c_j^{\phantom\dagger} + h.c. \right) 
+ V \sum_j n_j^{\phantom\dagger} n_{j+1}^{\phantom\dagger} \; , 
\label{eq:hamiltonian}
\end{equation}
with nearest-neighbor hopping amplitude $t_{\rm h}$ and
nearest-neighbor interaction strength $V$ at half-filling. 
The $c_i^{\left( \dagger \right)}$ annihilate (create)
fermions on lattice site $i$, $n_i = c_i^{\dagger}
c_i^{\phantom\dagger}$, and we take $\hbar = 1$.
We measure energies in units of $t_{\rm h}$, and, accordingly, time.  
The well-known ground-state phase diagram for the half-filled system
consists of a Luttinger liquid (LL) for $V<V_c =  2 $,
separated from a charge-density-wave (CDW) insulator ($V>V_c$) by a
quantum critical point $V_c$ \cite{spinless_fermions}. 
This model is integrable, with an exact solution via the Bethe
ansatz \cite{HBenergy}. 
We consider open chains of up to $L=100$ sites 
pushed out of equilibrium by suddenly quenching the strength of $V$
from an initial value $V(t=0)=V_0$ to a different value $V(t>0) = V$. 
Furthermore, we study the effect of adding a next-nearest-neighbor
repulsion $V_2 \sum_j n_j^{\phantom\dagger} n_{j+2}^{\phantom\dagger}$
to the model, which makes it non-integrable.
We compute the time evolution using the Lanczos time-evolution
method \cite{lubich_timeevolve,manmana:269,noack:93,ED_review} and the
adaptive t-DMRG.
We study the momentum distribution function (MDF)
$
\langle n_k \rangle(t) = \frac{1}{L} \sum_{l,m=1}^L e^{i k
  (l-m)} \langle c^{\dagger}_l c_m \rangle(t),
$
i.e., the Fourier transform of the one-particle density matrix, 
$\rho_{lm} = \langle c^{\dagger}_l c_m \rangle$.
In the t-DMRG, we utilize the Trotter approach developed in 
Refs.\  \cite{Uli_timeevolve,white:076401} as well as the
Lanczos approach \cite{schmitteckert:121302,manmana:269} with additional
intermediate time steps added within each time interval \cite{feiguin:020404}.
We hold the discarded weight fixed to 
$\varepsilon \leq 10^{-9}$ during the time evolution, but additionally
restrict the number of states kept to be in the range $100 \leq m \leq 1500$.
In all calculations presented here, the maximum error in the
energy, which is a constant of motion, is 1\%, and, in most cases, less
than 0.1\%, at the largest times reached.

\begin{figure}[hb]
\includegraphics[width=0.31\textwidth]{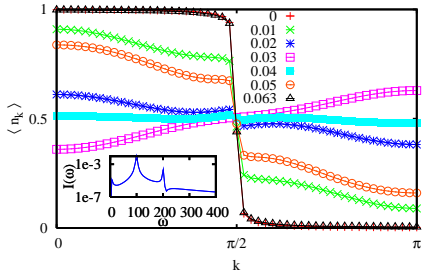}
 \caption{(color online) Time evolution of an initial LL state ($V_0 = 0.5$) in
   the strong coupling limit with $V = 100$ 
   for $L = 100$ sites at the times indicated. 
   Inset:
   spectral analysis of $\langle n_\pi(t) \rangle$.
} 
 \label{fig::strongcopuling_limit}
 \end{figure}

When an initial LL state is quenched to the strong-coupling regime,  
$V \gg t_{\rm h}$,
we find that $\langle n_k(t) \rangle$
(Fig.\ \ref{fig::strongcopuling_limit}), exhibits collapse and
revival on short time scales, whereas the 
density-density correlation function remains essentially unchanged,
i.e., retains the power-law decay of the LL. 
This can be understood by considering 
a quench to the atomic limit, $t_{\rm h} = 0$.
In this limit, all observables that commute with the density operator,
including the density-density correlation function, are
time-independent.
Furthermore, since the only remaining interaction is the
nearest-neighbor density-density interaction, it can be shown
analytically that the one-particle
density matrix $\rho_{m\,l}(t)$ involves only two frequencies
($\omega_1= V$ and $\omega_2=2V$), resulting in a periodic oscillation
with a revival time of $T_{\rm revival} = 2 \pi/V$ \cite{UnpublishedSalva}.  
Thus, in analogy to the observed collapse and revival of a BEC in an
optical lattice \cite{Nature:Bloch}, 
the single-particle properties of an initial LL state 
exhibit collapse and revival with this period. 
For the strong-coupling regime,
the time evolution
retains the oscillatory behavior of the atomic limit; 
two frequencies $\omega_1$ and
$\omega_2$ are indeed dominant in the spectrum, as can be seen in the
inset of  Fig.\ \ref{fig::strongcopuling_limit}. 
However, the finite hopping amplitude leads to a dephasing of the
oscillation on a time scale of $t_{\rm dephase} \sim 1/t_h$. 

\begin{figure}[hb]
\includegraphics[width=6.95cm]{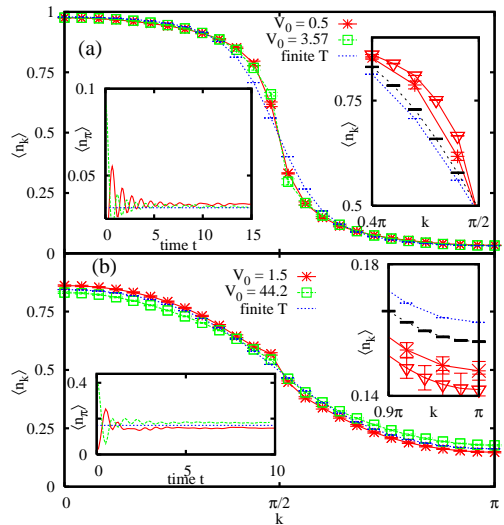}
\caption{(color online) Time-averaged momentum distributions when
  quenching (a) 
  from $V_0 = 0.5$ to $V=2$ %NEW
  (quantum critical point) and (b) 
  from $V_0 = 1.5$ to $V=5$ 
(insulator) 
for $L=50$
  sites. 
  The time averages of two independent initial states with the same
  energy are compared to each other and to the thermal expectation
  value.
  In the right inset, results for $L=50$ (\textcolor{red}{$\ast$}) are
  compared to $L=100$ (\textcolor{red}{$\bigtriangledown$}) for
  the regions with the largest differences.
  As a reference, finite $T$ data for $L=50$ (dotted line) and $L=100$
  (dashed line) are shown. Left insets: $\langle n_\pi \rangle$
  vs. time t; the horizontal line is the finite $T$ value.}
\label{fig:small_changes}
\end{figure}

Afterwards, observables oscillate with a small amplitude around a
fixed value, suggesting that the system reaches 
a quasi-stationary state.
In order to further characterize such states, we study the evolution
of the system 
for various values of $V$
%NEW SRM:
up to times one order of magnitude larger than $1/t_h$ when applying the
t-DMRG and up to two orders of magnitude larger when using full
diagonalization (FD). 
We find that the time averages for the longer times reachable by the
FD agree with the time averages for the times reachable by the
t-DMRG. 
Therefore, we conclude that the relevant time scale for the relaxation
is indeed given by $1/t_h$.
In Fig.\ \ref{fig:small_changes}, the MDFs, obtained by
performing an average in time from time $t = 3$ to $t = 10$ at the
quantum critical point, $V=V_c$, and at a point in the CDW region,
$V=5$, are shown. 
In order to investigate to what extent the
(quasi-)stationary behavior is generic, we examine its dependence on
the initial state.
We do this by preparing two qualitatively different initial states with the
same average energy $\langle \hat H \rangle$ for each case: 
one a ground state in the LL regime
and the other a ground state in the CDW regime.
This is possible for a certain range of $V$ in the intermediate
coupling regime.
In Fig.\ \ref{fig:small_changes}, results for two such initial states are
compared with each other and with the MDF obtained for a system in
thermal equilibrium and the same average energy, calculated using QMC
simulations \cite{QMC}.

At the critical point, $V=V_c$, Fig.\ \ref{fig:small_changes}(a), 
the MDFs for the two initial states coincide with each
other, to within the accuracy of the calculations (approximately the
symbol size) or less.
Therefore, information about the initial state is not preserved in this
quantity, consistent with the expectation for an ergodic evolution. 
However, the difference from the thermal distribution is
significant; thermalization is not attained.
The left inset shows the time evolution of
$\langle n_k \rangle (t)$ for $k = \pi$ for both initial conditions,
demonstrating that the system reaches a quasi-stationary state.
A small, but discernible shift from the thermal value (horizontal
line) can also be seen for both initial conditions, even at $k=\pi$. 
In the right inset, the points with the largest differences just below
the Fermi vector $k_F=\pi/2$ are plotted when the system size is
doubled from $L=50$ to $L=100$.
%NEW SRM:
%As can be seen, increasing the system sizes does not alter the
%conclusions.
An analysis of the data for $L=100$ does not alter our conclusion. 

In order to investigate the importance of quantum criticality,
we have also examined the behavior for $V=1.5$ and $V=2.5$ (not
shown), i.e., below and above the transition point.
We find almost identical behavior, indicating that the lack of
thermalization is not associated with the quantum critical point.
Note, however, that the LL regime ($V < 2$) is, in a sense,
generically critical.

\begin{figure}
\includegraphics[width=0.3\textwidth]{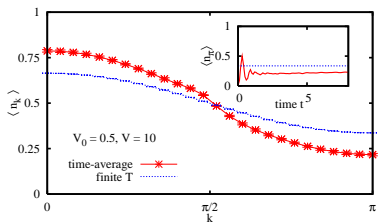}
\caption{(color online)
Time-averaged momentum distribution $\langle n_k \rangle$ for V=10.
Inset: $\langle n_\pi\rangle$ vs.\ time $t$; the horizontal line
is the finite-$T$ value. 
\label{fig:largeV}}
\end{figure}

For $V=5$, 
Fig.\ \ref{fig:small_changes} (b), all three
curves show small, but significant differences.
This means that the time evolution starting from different
initial states can be distinguished from each
other as well as from the thermal state, i.e., neither
relaxation to one distinguished quasi-stationary state 
nor thermalization occurs.
For this case, $\mid V_0 -V\mid$ is larger than for $V=2$,
and, in addition, the values of $V_0$ necessary to obtain the same energy in
the initial state ($V_0 = 1.5$ and $V_0 = 44.2165$) differ strongly.
This suggests that the initial states are far apart from 
each other in some sense, a notion that will be made more precise below.

The differences with the thermal 
distribution increase for larger $\mid V_0 - V\mid$.
As can be seen in Fig.\ \ref{fig:largeV} for $V=10$, 
the difference between the time average and the thermal distribution
is significant, clearly
confirming that thermalization does not occur.
The differences observed increase gradually as a function of
$\mid V_0 - V\mid$,
ruling out a transition as suggested for the Bose-Hubbard model
\cite{corinna:quench}.

\begin{figure}[ht]
\includegraphics[width=0.3\textwidth]{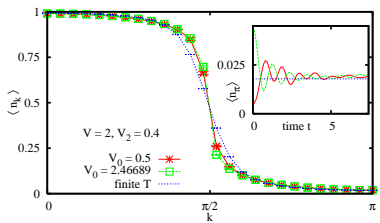}
\caption{(color online) 
Comparison of QMC and time-averaged values of the momentum
distribution $\langle n_k\rangle$
for the non-integrable case with $V=2$ and $V_2=0.4$.
Inset: the same as in Fig.\ \ref{fig:largeV}. 
}
\label{fig:nonintegrable}
\end{figure}

In order to investigate the impact of the lack of integrability on
thermalization, we now extend our model (\ref{eq:hamiltonian}) by
turning on a next-nearest-neighbor interaction. 
In Fig.\ \ref{fig:nonintegrable}, we display results with
$V_0=0.5$ and $V_0=2.46689$ (zero n.n.n.\ interaction), and the quenched
evolution at $V=2$, $V_2=0.4$. 
As in the integrable case, both initial states lead to indistinguishable 
time-averaged MDFs, but ones that are significantly
different from the thermal one, showing differences very similar to
those in Fig.\ \ref{fig:small_changes}(a).   
When $V_0=0.5$ and $V=10, \, V_2 = 1$ (not shown) the difference from
the thermal state is comparable to the case shown in
Fig. \ref{fig:largeV}. 
Therefore, the non-thermal nature of the emerging steady state is
clearly \textit{not} related to the integrability of the system. 

In order to shed light on the numerical results presented above and to 
characterize the quasi-stationary state, we consider a generalized
ensemble in which the expectation values of higher moments of $\hat
H$, which are constants of the motion, are taken as constraints. 
Note that the usual thermal density matrix $\hat{\varrho}_\beta$
is uniquely fixed by the single constraint
$\langle \hat H \rangle_\beta=\langle \hat H \rangle$.
Rigol {\em et al}.\ \cite{marcos:relax},
find a generalized Gibbs ensemble, in which
the density matrix is determined by maximizing the entropy taking into
account the constraints, to be an appropriate 
choice \cite{marcos:relax, balian,marcos:relax2}. 
The general form of the statistical operator is then 
$
\hat{\varrho} = \exp\left[-\sum_n \lambda_n \hat{O}_n\right],
$
where 
the operators $\hat{O}_n$ form a set of observables whose expectation
values remain constant in time.
The values of the $\lambda_n$ 
are fixed by the condition that ${\rm Tr} \left( \hat{\varrho}\,
  \hat{O}_n \right) =\langle \hat{O}_n \rangle$, with $\hat{O}_0 = 1$
to enforce normalization. 
In some special cases like hard-core bosons in one dimension
\cite{marcos:relax,marcos:relax2} or the Luttinger model
\cite{cazalilla:156403},
constants of motion can be found in terms of operators in second quantization.
However, this is not possible for Bethe-ansatz-integrable systems.
For any {\it closed} system, however, the quantities
$\hat{O}_n = {\hat H}^n$ can be used.
Taking all powers as constraints would unambiguously fix all correlation
functions to all lengths.
For a finite system,
it can be shown that
$\hat{\varrho}$ is fully determined by
$\text{dim}(\hat{H})$ powers of $\hat H$
\cite{UnpublishedSalva}, 
%NEW SRM:
for $\hat H$ with a bounded spectrum.  
The statistical expectation
value of any observable is then given by ${\rm
  Tr}\left(\hat{\varrho} \: \hat{O} \right) = \sum\limits_\nu |\langle
\nu | \psi_0 \rangle|^2 \langle \nu | \hat{O} | \nu \rangle$ (for a
non-degenerate spectrum),
%NEW SRM:
where $|\psi_0 \rangle$ is the initial state and $|\nu \rangle$ are
the eigenstates of $\hat{H}$.
It can be easily seen, that the
r.h.s.\ of this expression equals the time average of $\langle \hat{O}
\rangle(t)$. 

\begin{figure}
\begin{center}
\includegraphics[width=0.3\textwidth]{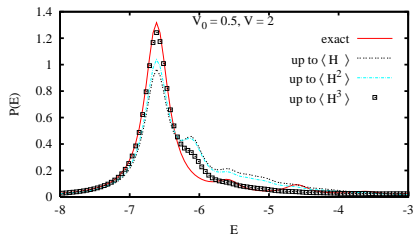}
\end{center}
\caption{(color online) Influence of the constraints $\langle
  \hat{H}^n \rangle$ on the energy distribution function $P(E)$. 
}
\label{PofE}
\end{figure}

We now investigate the extent to which
the statistical expectation value within the generalized Gibbs ensemble 
approaches the time average of the evolution after a quench by
studying the energy distribution
for a given state $|\psi\rangle$%NEW
, defined as 
$P_\psi(E) = \sum_\nu \delta \left(E-\epsilon_\nu\right) 
|\langle \nu | \psi \rangle|^2
$,
which is normalized if $\langle \psi | \psi \rangle= 1$. 
The energy distribution in the 
generalized Gibbs ensemble can analogously be defined as
$
P_G(E) = {\rm Tr} \, \delta \left(E- \hat H \right) \hat{\varrho}
$,
with $\hat{\varrho}$ as defined previously. 
In Fig.\ \ref{PofE}, we show $P_\psi (E)$ calculated using full
diagonalization on $L=16$ sites for an initial state
$V_0=0.5$ evolved at the quantum critical point, $V=2$ compared with 
the distribution in the Gibbs ensemble $P_G (E)$ as the number of
constraints is increased from 1 to 3.
It is evident that increasing the number of constraints
systematically improves the agreement and
that only a small number of
moments are necessary to obtain very good agreement.

The distance between two distributions can be estimated using the
moments of the absolute differences,
$\Delta_n = \int {\rm d}E \, E^n \, \mid P(E) - P'(E) \mid \,$.
Taking $\Delta_0/W$, with $W$ the bandwidth of $\hat{H}$,
as an estimate of $\mid P(E) - P'(E) \mid$,  
we see that  
the difference between moments of $\hat H$ for two different
 energy distributions $P$ and $P'$ can be estimated as 
 \begin{equation}
 \langle {\hat H}^n \rangle_P - \langle {\hat H}^n \rangle_{P'}
  \leq  \Delta_n \simeq \frac{1}{n+1} W^n \Delta_0 \; .
 \end{equation} 
Therefore, if the distance between the distributions $\Delta_0 \ll 1$,
then the relative difference of the moments $\left( \langle {\hat H}^n
  \rangle_P - \langle {\hat H}^n \rangle_{P'} \right)/W^n <
\Delta_0/(n+1)$ will also remain small, and observables will converge to
values close to each other after a quench. 
For the cases of evolution with metallic and insulating initial states
discussed above, we obtain  
$\Delta_0 = 0.12439$ for $V_0=0.5$ and $V_0=3.57463$ ($V=2$), and 
$\Delta_0 =0.41521$ for $V_0 =1.5$ and $V_0=44.2165$ ($V=5$).
On the other hand, comparison of $P_\psi$ with the thermal
distribution $P_{\beta}$ yields  
$\Delta_0 = 0.68581$ ($V_0=0.5$, $V=2$), and $\Delta_0 = 1.24616$
($V_0=1.5$, $V=5$), respectively.
Thus, the distance between the thermal distribution and the one defined 
by the initial states is always larger than those defined by the pair
of initial states with the same energy, 
%RMN in agreement with the fact that 
%thermalization was not observed.
supporting our observation that thermalization does not occur.

% NEW SRM:
In summary, our adaptive time-dependent density-matrix renormalization
group simulations of the time evolution of a system of correlated
spinless fermions after a quantum quench have exhibited the following
generic behavior: Independently of its integrability or criticality,
the system relaxes to a non-thermal quasi-stationary state. 
Observables relax to the same value when different initial states have
the same energy and are sufficiently close to each other, i.e., the
memory of the initial state is lost in the observables after
relaxation. 
`Closeness' can be quantified using a measure $\Delta_0$ which is
based on the energy distributions defined for the initial state or for
a given density matrix.  
Increasing the number of constraints (moments of $\hat H$) in a
generalized Gibbs ensemble leads to convergence to the energy
distribution defined by the initial state. 

We acknowledge useful discussions with M.\ Arikawa, F.\ Gebhard,
C.\ Kollath, S.R.\ White, and especially M.\ Rigol. 
S.R.M. acknowledges financial support by SFB 382 and SFB/TR 21.
We acknowledge HLRS Stuttgart and NIC J\"ulich for allocation of CPU
Time.

\end{document}